\documentclass{article}

\usepackage[final, nonatbib]{nips_2018}

\usepackage[utf8]{inputenc} 
\usepackage[T1]{fontenc}    
\usepackage{amsmath,amssymb}
\usepackage[pdftex]{graphicx}
\usepackage{hyperref}       
\usepackage{url}            
\usepackage{booktabs}       
\usepackage{nicefrac}       
\usepackage{microtype}      

\title{Link Prediction in Dynamic Graphs for Recommendation}

\author{
  Samuel G.~Fadel \\ 
  Institute of Computing \\
  University of Campinas \\
  Campinas, Brazil \\
  \texttt{samuel.fadel@ic.unicamp.br} \\
  \And
  Ricardo da S.~Torres \\
  Institute of Computing \\
  University of Campinas \\
  Campinas, Brazil \\
  \texttt{rtorres@ic.unicamp.br}
}

\DeclareRobustCommand{\spref}[3]{#2} 
\begin{document}

\maketitle

\begin{abstract}
Recent advances in employing neural networks on graph domains helped push the state of the art in link prediction tasks, particularly in recommendation services.
However, the use of temporal contextual information, often modeled as dynamic graphs that encode the evolution of user-item relationships over time, has been overlooked in link prediction problems.
In this paper, we consider the hypothesis that leveraging such information enables models to make better predictions, proposing a new neural network approach for this.
Our experiments, performed on the widely used ML-100k and ML-1M datasets, show that our approach produces better predictions in scenarios where the pattern of user-item relationships change over time.
In addition, they suggest that existing approaches are significantly impacted by those changes.
\end{abstract}

\section{Introduction}
In recommendation systems, users are offered a set of items they might be potentially interested in.
One common approach to this problem is collaborative filtering~\cite{goldberg_using_1992}, where the collective ratings users give to items are used to predict unseen ratings.
This is commonly represented as an undirected bipartite graph, where observed ratings are weighted edges linking one user to one item, with the weight representing the rating value.
Following recent advances on recommender systems using neural networks and inspired by collaborative filtering~\cite{berg_graph_2017,monti_geometric_2017}, we also approach this problem as a link prediction task.
The goal in link prediction is to learn, from an observed set of edges, a model that predicts the weight of unseen edges.
Here, predicting this weight is equivalent to predicting what rating the user would give to an item, thus estimating how likely they are to be interested in it.

One particular issue with existing approaches, however, is that they do not take into account that users might change their perception of items over time.
Would the use of temporal contextual information improve link prediction results in recommendation tasks?
In this paper, we address this research question by modeling temporal contextual information related to user-item relationships as dynamic graphs.
Our hypothesis is that we could leverage this information to produce prediction models that take advantage of how users change their tastes, or even how some item is perceived by other users, over time.
This enables new types of collective information about them to arise, such as when some movie is relevant only for a short period of time or users discover some new genre of music they are now increasingly interested in.
Our solution builds upon Graph Convolutional Matrix Completion (GCMC)~\cite{berg_graph_2017} to incorporate dynamic graphs.
Experiments performed on the widely used ML-100k and ML-1M movie recommendation datasets suggest that the use of dynamic graphs lead to better link prediction results than those observed for state-of-the-art baselines.

\section{Recommendation as Link Prediction in a Dynamic Graph}
We are interested in investigating how the temporal information can be leveraged, building upon existing approaches that pose the task of recommending items to users as a link prediction task.
To do this, we extend GCMC~\cite{berg_graph_2017}, a method for matrix completion using Graph Convolutional Networks (GCNs), to handle dynamic graphs.

Formally, let $\mathcal{U} = \{ u_i | 1 \leq i \leq n_u \}$ and $\mathcal{V} = \{ v_j | 1 \leq j \leq n_v \}$ denote the set of users and items, respectively, with $n_u$ being the number of users and $n_v$ being the number of items.
Also let $G = (\mathcal{W}, \mathcal{E}, \mathcal{R})$ be the ratings bipartite graph, where $\mathcal{W} = \mathcal{U} \cup \mathcal{V}$ is the set of nodes, $\mathcal{E}$ is the set of edges, representing all observed ratings, and $\mathcal{R}$ is the set of possible rating values (edge types).
Each observed rating is denoted by $(u_i, r, v_j) \in \mathcal{E}$, with $u_i \in \mathcal{U}$, $v_j \in \mathcal{V}$, and $r \in \mathcal{R}$.
This graph is represented by an adjacency matrix $M$, where $M_{ij} = r$ corresponds to the edge $(u_i, r, v_j)$.
We also consider a set of feature vectors for users and items, denoted as $\mathcal{X} = \{ \mathbf{x}_w | w \in \mathcal{W} \}$.

\subsection{Graph convolutional matrix completion (GCMC)}
Put simply, GCMC predicts ratings similarly to an autoencoder: it first encodes users and items, then uses a decoder that takes a pair of encoded user $\mathbf{z}_{u_i}$ and item $\mathbf{z}_{v_j}$ to predict the rating the user $u_i$ would provide to the item  $v_j$.
In GCMC, the encoding step for each user is comprised of a GCN-like graph convolution step followed by a fully connected layer:
\begin{equation}\label{eq:gcmc-hidden}
    \mathbf{h}_{u_i} = \xi \left(
               \mathrm{accum}_{r \in \mathcal{R}}\left\{
                   \sum_{j \in \mathcal{N}_{i, r}} \frac{1}{c_{ij}} W_r \mathbf{x}_{v_j}
               \right\}
           \right),
\end{equation}
\begin{equation}\label{eq:gcmc-output}
    \mathbf{z}_{u_i} = \xi(W \mathbf{h}_{u_i}),
\end{equation}
where $N_{i, r}$ is the set of indices $j$ such that $M_{ij} = r$, $c_{ij}$ is a normalization constant, $\mathrm{accum}(\cdot)$ denotes an accumulation function that can be either a vector concatenation or sum operation, $\xi(\cdot)$ is an activation function, both $W_r \in \mathbb{R}^{d_\mathrm{hidden} \times d_\mathrm{input}}$ and $W \in \mathbb{R}^{d_\mathrm{output} \times d_\mathrm{hidden}}$ are learned during training, and $d_\textrm{input}$, $d_\textrm{hidden}$, and $d_\textrm{output}$ are the dimensionalities of the input features ($\mathbf{x}$ vectors), encoder hidden layer  (Equation~\ref{eq:gcmc-hidden}) and encoder output layer (Equation~\ref{eq:gcmc-output}), respectively.
Note that there is one binary indicator adjacency matrix for each type of edge in $\mathcal{R}$.
Each item $v_j$ is also encoded in an analogous way, possibly sharing parameters $W_r$ and $W$ with the user encoder.

The ratings are then predicted using one bilinear decoder per rating value, followed by a softmax operation.
The predicted probability of the rating from user $u_i$ with respect to the item $v_j$ being $r$ is:
\begin{equation}\label{eq:gcmc-decoder}
    P(\hat{M}_{ij} = r) = \frac{\exp \left( \mathbf{z}_{u_i}^T Q_r \mathbf{z}_{v_j} \right)}
                               {\sum_{s \in \mathcal{R}} \exp \left(\mathbf{z}_{u_i}^T Q_s \mathbf{z}_{v_j} \right)},
\end{equation}
where each $Q_r \in \mathbb{R}^{d_\mathrm{output} \times d_\mathrm{output}}$, $r \in \mathcal{R}$, is learned during training.
With both the encoder and decoder defined, we train the model to minimize the negative log likelihood of the predicted ratings:
\begin{equation}
    \mathcal{L}(M, \hat{M}) = -\sum_{(u_i, r, v_j) \in \mathcal{E}} \log P(\hat{M}_{ij} = r).
\end{equation}

\subsection{Leveraging Temporal Information}
Under the framework presented, we view ratings as new edges being added to a graph that changes over time, with future edges representing the unknown ratings we want to predict.
Additionally, since this is used for recommending items, we are not interested in predicting \emph{when} some rating will take place, only its \emph{value}.
The dynamic graph is a sequence of graphs $G^{(1)}, G^{(2)}, \ldots, G^{(T)}$, with $T > 1$, representing states over time and is such that it contains all observed ratings.
Equivalently, this can be viewed as a series of rating matrices $M^{(1)}, M^{(2)}, \ldots, M^{(T)}$.

We consider two possible representations for this.
For both, input sequences have a fixed size $T$, independent of the total number of ratings.
In the first strategy, the dynamic graph is thought of as a sliding ``window'' over the series of ratings, and we represent steps in the sequence as a series of disjoint sets of edges where each $M^{(t)}$ contains approximately the same number of edges.
In contrast, the second strategy adopts a more explicit representation of rating history, as each $M^{(t+1)}$ contains the new ratings observed so far, plus all ratings observed in the previous time step $M^{(t)}$, with approximately the same number of ratings added at each step.

\begin{figure}[t]
    \centering
    \includegraphics[width=\textwidth]{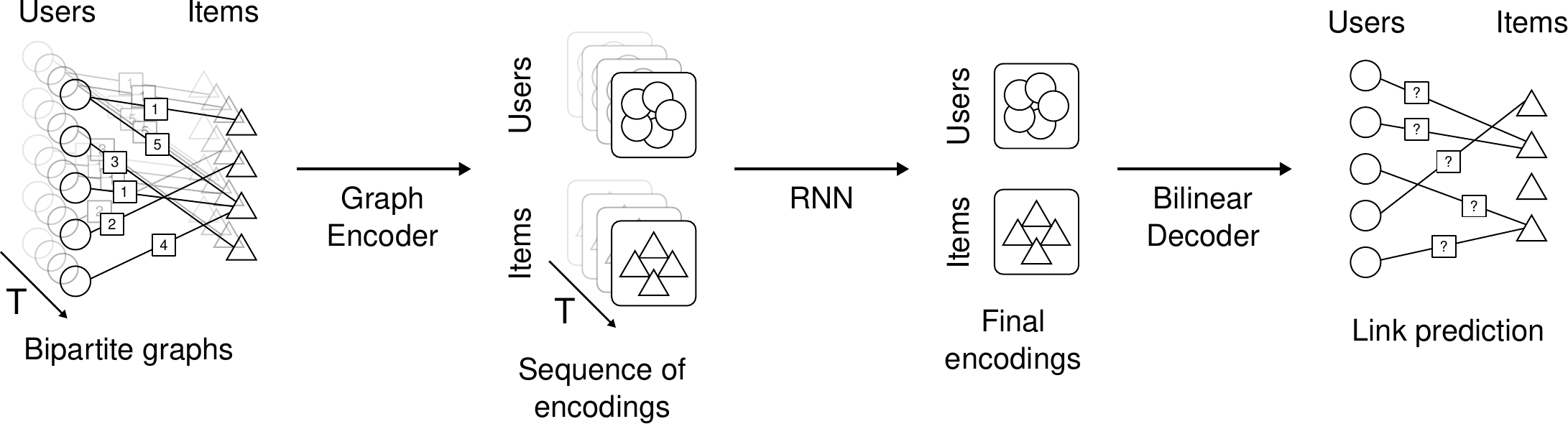}
    \caption{
        Our pipeline for temporal link prediction, based on previous approaches~\cite{berg_graph_2017}.
        The model uses $T$ matrices representing ratings as they happen over time.
        Those can be either incremental, that is, every new rating is incorporated in the next step, or disjoint, where each rating matrix has approximately the same number of ratings.
    }
    \label{fig:gcmc-seq}
\end{figure}

In order to handle the sequence of matrices, we introduce a new step in the encoding process.
First, each $M^{(t)}$ is encoded using the same (shared) encoder (Equations \ref{eq:gcmc-hidden} and \ref{eq:gcmc-output}), producing the user $\mathbf{z}_{u_i}^{(t)}$ and item $\mathbf{z}_{v_j}^{(t)}$ encodings for each time step $t$.
Then, a recurrent layer with $d_\mathrm{hidden}$ units maps the encodings sequence $\mathbf{z}_{u_i}^{(t)}$ (and $\mathbf{z}_{v_j}^{(t)}$, respectively) into a single encoding $\mathbf{z}_{u_i}$ (and $\mathbf{z}_{v_j}$, respectively), which is the final hidden state of the recurrent layer.
Finally, these can be decoded into ratings with the same decoder used by GCMC (Equation~\ref{eq:gcmc-decoder}).
An overview of this approach is illustrated in \autoref{fig:gcmc-seq}.

\section{Experimental setup}
Following previous results~\cite{berg_graph_2017}, we use the same number of parameters for both the encoder and decoder, namely $d_\mathrm{hidden} = 500$ and $d_\mathrm{output} = 75$, and the same weight-sharing scheme for the $W_r$ and $Q_r$ parameters.
The $W$, $W_r$, $Q_r$, and the recurrent layers share the same weights for encoding both users and items.
We also employ dropout ($p = 0.7$) on the input of the graph encoder~(Equation~\ref{eq:gcmc-hidden}) and on the resulting encodings (Equation~\ref{eq:gcmc-output}).
Activation functions used are rectified linear units (ReLUs), except for the recurrent layer.
All models were trained using Adam~\cite{kingma_adam_2015} using a learning rate of $10^{-2}$.
We performed experiments with both LSTM~\cite{hochreiter_long_1997} and GRU~\cite{cho_learning_2014} layers, using $T = 10$.

\paragraph{Datasets}
We performed experiments on the widely known ML-100k~\cite{harper_movielens_2015} and ML-1M~\cite{harper_movielens_2015} datasets.
While there is evidence that using features for users or items can improve recommendations~\cite{berg_graph_2017}, we avoid using them in order to minimize the influence of factors other than the additional temporal information, as the goal of these experiments lies in highlighting how it can be leveraged to improve recommendations.
Hence, $\mathcal{X}$ is a set of 1-hot encoded vectors to represent each node in the graph, with $d_{\mathrm{input}} = |\mathcal{W}|$.

In contrast to previous work, we adopt different training and test splits in our comparisons, while still including the original splits for comparison.
As the test set of both datasets contains ratings that happened before some from the training set, we cannot properly evaluate our models since we cannot use future ratings to predict those that happen in the past.
Consequently, we define new training, validation, and test sets where we guarantee that every rating in the training set happened before every rating in either the validation or test sets, thus avoiding this issue.
The test set comprises 20\% of the whole dataset (original train and test splits), the validation set is 20\% of the remaining data, while the rest is used for training.

In order to minimize divergences when comparing with the baseline~\cite{berg_graph_2017}, we also adopt the following for each dataset.
For ML-100k, we used vector concatenation as the $\mathrm{accum}(\cdot)$ operation and trained all models for $1000$ epochs.
For ML-1M, $\mathrm{accum}(\cdot)$ is a sum operation and all models are trained for $3500$ epochs.
Additionally, the constants $c_{ij}$ (Equation~\ref{eq:gcmc-hidden}) are computed using left normalization for ML-100k and symmetric normalization for ML-1M.

\paragraph{Evaluation}
We follow the same evaluation protocol adopted recently for this task~\cite{berg_graph_2017,monti_geometric_2017}.
The final predicted rating, defined as follows, is the expected value of the rating probabilities from Equation~\eqref{eq:gcmc-decoder}:
\begin{equation}\label{eq:predicted-rating}
    \mathbb{E}[\hat{M}_{ij}] = \sum_{r \in \mathcal{R}} r P(\hat{M}_{ij} = r).
\end{equation}
With these values, we compute the root mean square error (RMSE) between the predicted and observed ratings, thus better predictions result in lower RMSE.
Results are computed on models that use averaged parameters with an exponential decay of $0.995$.
We report RMSE values averaged over 5 runs with different parameter initializations.

\paragraph{Baselines}
We compare the performance of our models to the original GCMC~\cite{berg_graph_2017}.
Since GCMC attains results very close to or at the state of the art on both datasets, it allows us to properly evaluate our approach with a strong baseline.
We note that while other methods, such as MGCNN~\cite{monti_geometric_2017}, could be used in our comparisons, their results are very similar to GCMC, and do not add significant variety to the analysis.

\section{Results and discussion}
All results are summarized in \autoref{tab:recommendation-rmse}.
At first glance, the most striking result is the difference in performance GCMC displays when the training and test sets were changed to the new ones.
As the original training and test sets are closer to a uniform sample taken from the whole dataset, their distributions are similar.
Once we view this from a temporal perspective, that is, unseen data is comprised of future ratings, it is evident that such a change in the distribution severely affects GCMC.
This highlights the importance of taking into account the evolution of user-item relationships happening over time.

\begin{table}
    \caption{
        RMSE values for all models on the ML-100k and ML-1M datasets using the new training and test sets, averaged over 5 different runs, along with standard deviations.
        GCMC (original) results are the same as those reported in previous work~\cite{berg_graph_2017}.
        The $\dagger$ symbol denotes experiments using disjoint edge sets in every $M^{(i)}$; $\ddagger$ denotes the representation where every step contains all edges from previous steps.
    }
    \label{tab:recommendation-rmse}
    \centering
    \begin{tabular}{lrrr}
        \toprule
        Method               & ML-100k             & ML-1M  \\
        \midrule
        GCMC (original)      & $0.910$             & $0.832$ \\
        GCMC (new split)     & $1.2259 \pm 0.0001$ & $1.1414 \pm 0.0002$ \\
        GCMC-LSTM$^\dagger$  & $1.2556 \pm 0.0509$ & $1.1857 \pm 0.0193$ \\
        GCMC-GRU$^\dagger$   & $1.2446 \pm 0.0404$ & $1.2056 \pm 0.0171$ \\
        GCMC-LSTM$^\ddagger$ & $1.0725 \pm 0.0217$ & $\mathbf{1.0394} \pm 0.0230$ \\
        GCMC-GRU$^\ddagger$  & $\mathbf{1.0630} \pm 0.0141$ & $1.0407 \pm 0.0336$ \\
        \bottomrule
  \end{tabular}
\end{table}

With this in mind, we see that the recurrent models attain better results than the baseline only in one of the representations.
While both LSTM and GRU variants displayed very similar performance, the disjoint edge representation performs worse than its alternative.
The causes for this might lie in the fact that the model cannot represent the entire set of edges after processing the whole sequence.

As the recurrent models have the exact same configuration in both scenarios, the representation is clearly the main factor influencing the performance of the model.
When using all edges from previous steps at every step, the performance gap is evident.
What this suggests is that being explicitly shown the complete set of available edges at each time step, the model can better generalize how the evolution of user-item relationships takes place in the future.

\subsubsection*{Acknowledgments}

Authors are grateful to CAPES (grant \#88881.145912/2017-01), CNPq (grant \#307560/2016-3), FAPESP (grants \#2017/24005-2, \#2014/12236-1, \#2015/24494-8, \#2016/50250-1, and \#2017/20945-0), and the FAPESP-Microsoft Virtual Institute (grants \#2013/50155-0, \#2013/50169-1, and \#2014/50715-9). This study was financed in part by the Coordena\c c\~ao de Aperfei\c coamento de Pessoal de N\'ivel Superior - Brasil (CAPES) - Finance Code 001.


\small
\DeclareRobustCommand{\spref}[3]{#3} 
\bibliography{paper}


%

\end{document}